\documentclass[aps,prl,twocolumn,showpacs,preprintnumbers,superscriptaddress,nofootinbib]{revtex4-1}
\usepackage{graphics}  
\usepackage{epsfig}
 \usepackage{verbatim}
\usepackage{subfigure}

\IfFileExists{srcltx.sty}{\usepackage[active]{srcltx}}

\newcommand{\be}{\begin{equation}}
\newcommand{\ee}{\end{equation}}
\newcommand{\bea}{\begin{eqnarray}}
\newcommand{\eea}{\end{eqnarray}}

\begin{document}

\title{PeV neutrinos from intergalactic interactions of cosmic rays emitted \\ by active galactic nuclei}

\author{Oleg E. Kalashev}
\affiliation{Institute for Nuclear Research, 60th October Anniversary Prospect 7a, Moscow 117312 Russia}

\author{Alexander Kusenko}
\affiliation{Department of Physics and Astronomy, University of California, Los 
Angeles, CA 90095-1547, USA}
\affiliation{Kavli IPMU (WPI), University of Tokyo, Kashiwa, Chiba 277-8568, Japan}

\author{Warren Essey}
\affiliation{Department of Physics and Astronomy, University of California, Los 
Angeles, CA 90095-1547, USA}


\preprint{}

\begin{abstract}

The observed very high energy spectra of {\em distant} blazars are well described by secondary gamma rays produced in line-of-sight interactions of cosmic rays with background photons. 
In the absence of the cosmic-ray contribution, one would not expect to observe very hard spectra from distant sources,  but the cosmic ray interactions generate 
very high energy gamma rays relatively close to the observer, and they are not attenuated significantly.  The same interactions of cosmic rays are expected to produce a flux of neutrinos 
with energies peaked around 1~PeV.  We show that the diffuse isotropic neutrino background from many distant 
sources can be consistent with the neutrino events recently detected by the IceCube experiment.  We also find that the flux from any individual 
nearby source is insufficient to account for these events. The narrow spectrum around 1~PeV implies that some active galactic nuclei can accelerate protons to EeV energies.

\end{abstract}

\maketitle

The IceCube collaboration has detected two neutrinos with energies  $1.04 \pm 0.16$ and $1.14 \pm 0.17$\,PeV~\cite{icecube,Laha:2013lka}.  These neutrinos are either electron or tau neutrinos.  The muon analysis, currently under way, is expected to produce additional events (probably, with a lower energy resolution).  The narrow energy range in which the two neutrinos have been detected may be consistent with a  spectrum peaked in the PeV energy range, above the experimental threshold of 0.4 PeV and below the Glashow resonance that enhances detector sensitivity around 6.3~PeV~\cite{Bhattacharya:2011qu}. 
Only specific types of astrophysical sources can produce a peaked spectrum around a PeV~\cite{Cholis:2012kq}.  

Narrow spectra peaked around 1~PeV were predicted to arise from line-of-sight interactions of cosmic rays emitted by blazars~\cite{Essey:2009zg,Essey:2009ju,Essey:2010er}.  There is growing evidence that intergalactic cascades initiated by line-of-sight interactions of cosmic rays produced by active galactic nuclei (AGNs) are responsible for the highest-energy gamma rays observed from blazars~\cite{Essey:2009zg,Essey:2009ju,Essey:2010er,Essey:2011wv,Murase:2011cy,Razzaque:2011jc,Prosekin:2012ne,Aharonian:2012fu,Zheng}. 
As long as the intergalactic magnetic fields are in the femtogauss range~\cite{Essey:2010nd}, the spectra of distant blazars are explained remarkably well with secondary photons from such cascades~\cite{Essey:2009zg,Essey:2009ju,Essey:2010er}. In the absence of such contribution, some unusually hard intrinsic spectra~\cite{Stecker:2007zj,Lefa:2011xh,Dermer:2011uu} or hypothetical new particles~\cite{De_Angelis:2007dy} have been invoked to explain the data.   Models for hard intrinsic spectra of $\gamma$ rays can be constructed, but the natural ease with which secondary photons reproduce the data makes the explanation based on cosmic rays very appealing.  Furthermore, the lack of time variability of the most distant blazars at energies above TeV is in agreement with this hypothesis, which predicts that the shortest variability time scales for $z \gtrsim 0.15$ and $E\gtrsim 1\ {\rm TeV}$ should be greater than $(0.1-10^3)$~years, depending on the model parameters~\cite{Prosekin:2012ne}.

Proton acceleration in relativistic shocks is determined by the shock Lorentz factor, the magnetization of the pre-shock flow, and the orientation of the field relative to the shock propagation~\cite{Sironi:2010rb}. 
In AGN jets, the relative Lorentz factor between the pre-shock flow and 
the post-shock flow is not expected to be as high as in gamma-ray bursts (GRBs), which makes it difficult for AGNs to
achieve the proton energies as high as those in GRBs. 
Some exceptional conditions, such as a small angle of magnetic fields in the internal shocks, can enable an efficient acceleration of protons up to $E_{\rm p,max}\sim 10^8$~GeV~\cite{Sironi:2010rb}.\footnote{While AGN can be considered as candidate sources for cosmic rays of even higher energies, the origin of such ultrahigh-energy cosmic rays (UHECR) remains unclear.  The contributions of unusual supernova explosions, GRBs, and the possibility of UHECR nuclei from nearby sources remain viable possible explanations of UHECR with energies above $10^{18}$~eV~\cite{Gaisser:2013bla,UHECR}.}   It is likely that the distribution of AGNs is a decreasing function of $E_{\rm p,max}$, with the values $E_{\rm p,max} \gtrsim 10^8$~GeV still allowed, but uncommon.  The interactions of cosmic rays with extragalactic background light (EBL) produce neutrinos via the reaction $p\gamma_{\rm EBL}\rightarrow p \pi^+$, which has a sharp threshold around $E_{\rm th}\sim 10^8$~GeV (broadened by the energy distribution of the EBL 
photons).  
As long as the distribution of AGN with $E_{\rm p,max}$  decreases fast enough to make the contribution of CMB photons unimportant, most neutrinos are produced in interactions with EBL of the protons emitted 
by  AGNs with $E_{\rm p,max} \sim 10^8$~GeV.  The neutrino spectrum is, therefore, limited by the fraction $\sim (0.01 - 0.1)$ of the threshold energy from below and by $\sim (0.01 - 0.1)\times  E_{\rm p,max}$ from above, with a peak around the threshold energy (where more AGN contribute protons), $E_\nu \sim (0.01 - 0.1) \times 10^8\, {\rm GeV}\sim 1\,{\rm PeV}$.  

The mechanism thus predicts a peaked spectrum of neutrinos around 1~PeV.  We will examine whether these neutrinos can account for PeV neutrino events in IceCube. 

Assuming the scenario of Refs.~\cite{Essey:2009zg,Essey:2009ju,Essey:2010er,Essey:2011wv,Murase:2011cy,Razzaque:2011jc,Prosekin:2012ne,Aharonian:2012fu,Zheng}, we have considered two possibilities for the origin of  IceCube neutrinos: a single nearby source, and a combined contribution of distant sources.  Obviously, the gamma-ray background and the cosmic-ray spectrum should not exceed the observed fluxes.  We do not assume that the cosmic ray spectrum up to ultrahigh energies is explained by the same sources; as was pointed out in Ref.~\cite{Roulet:2012rv}, such a scenario disagrees with the data.  
Also, we do not consider neutrinos produced inside AGNs as in Ref.~\cite{Kistler:2013my}.  

For the case of one or a few nearby point sources, we have not been able to find an acceptable explanation of the IceCube events. 
Indeed, the neutrino required flux, $E_\nu^2 \frac{d{\cal F}}{dE_\nu} \sim 
 20  \ {\rm eV} \,  {\rm cm}^{-2} \, {\rm s}^{-1} \, {\rm sr}^{-1}$, is an order of magnitude greater than the predicted flux from a single source shown in Ref.~\cite{Essey:2009ju}. 
We have calculated numerically the expected number of events for the spectral shape of a blazar signal from  Ref.~\cite{Essey:2009ju} using the detector sensitivity plots available in Ref.~\cite{icecube}.  The sources mentioned in Ref.~\cite{Essey:2009ju} would not result in the observed numbers of events.  A single source with the same spectrum, but at a smaller distance from Earth would produce an unacceptably large flux of cosmic rays.    We proceed to considering the second possibility: a diffuse background from distant sources.

For the diffuse flux calculation we use the numerical code described in detail in Ref.~\cite{Gelmini:2011kg}. The code is based on kinetic equations; 
it calculates the propagation of nucleons, stable leptons  and photons using the standard dominant processes, {\em i.e.} pion production by nucleons, 
$e^{\pm}$ pair  production by protons and neutron $\beta$-decays. For electron-photon cascade development, it includes $e^{\pm}$ pair production and
inverse Compton scattering. We also take into account neutrino oscillations on their way from the site of production to the observer.  Since the distance traveled by neutrinos is much greater than the oscillation length, muon neutrinos oscillate into tau neutrinos with a 50\% probability.  The resulting spectrum has a flavor ratio of approximately (1:1:1).  Our numerical calculations we use the actual mixing angles in the tri-bimaximal neutrino mixing approximation.  

\begin{figure*}[ht!]
\subfigure[]{\includegraphics[angle=270,width=0.47\textwidth]{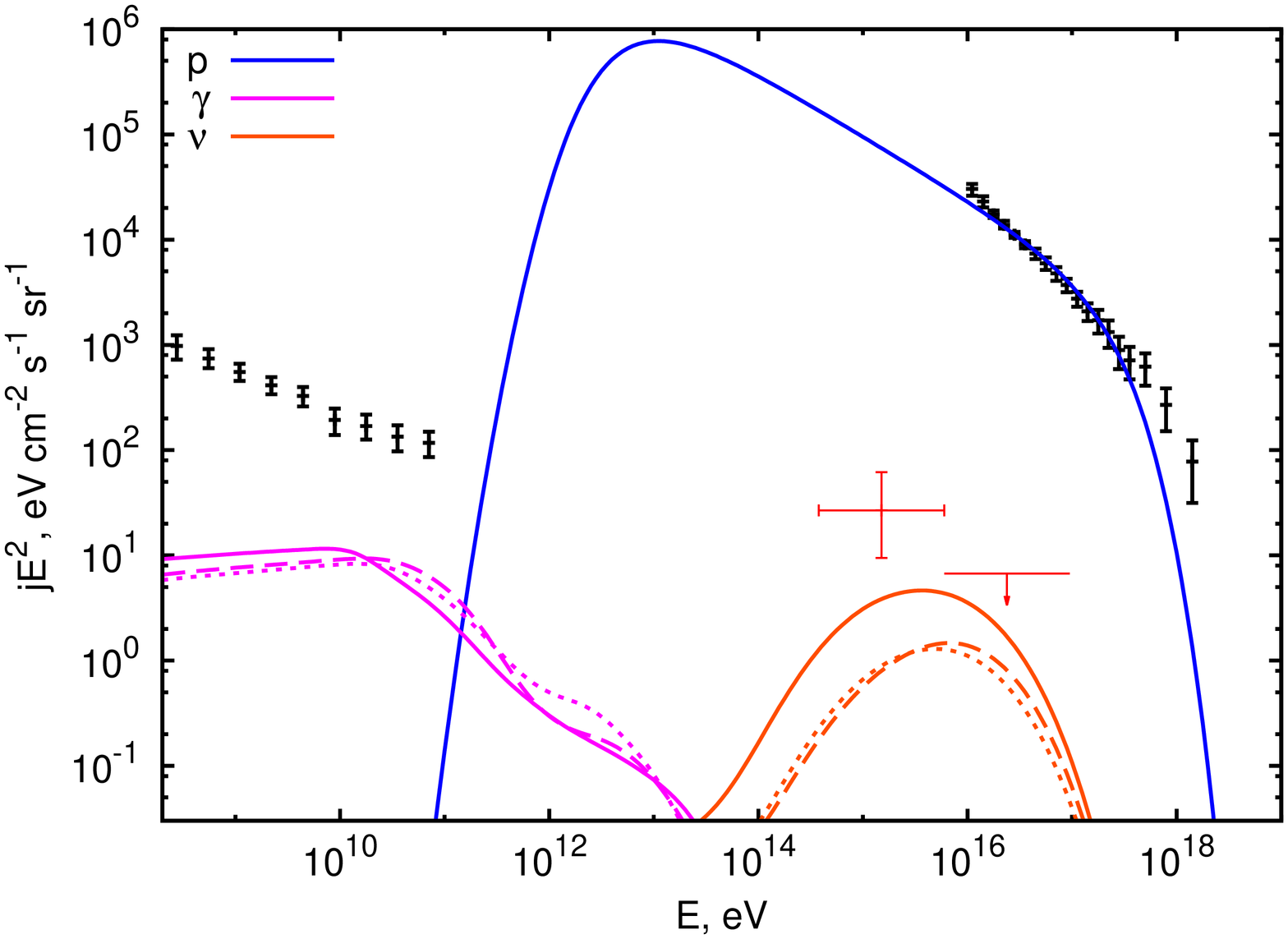}} 
\subfigure[]{\includegraphics[angle=270,width=0.47\textwidth]{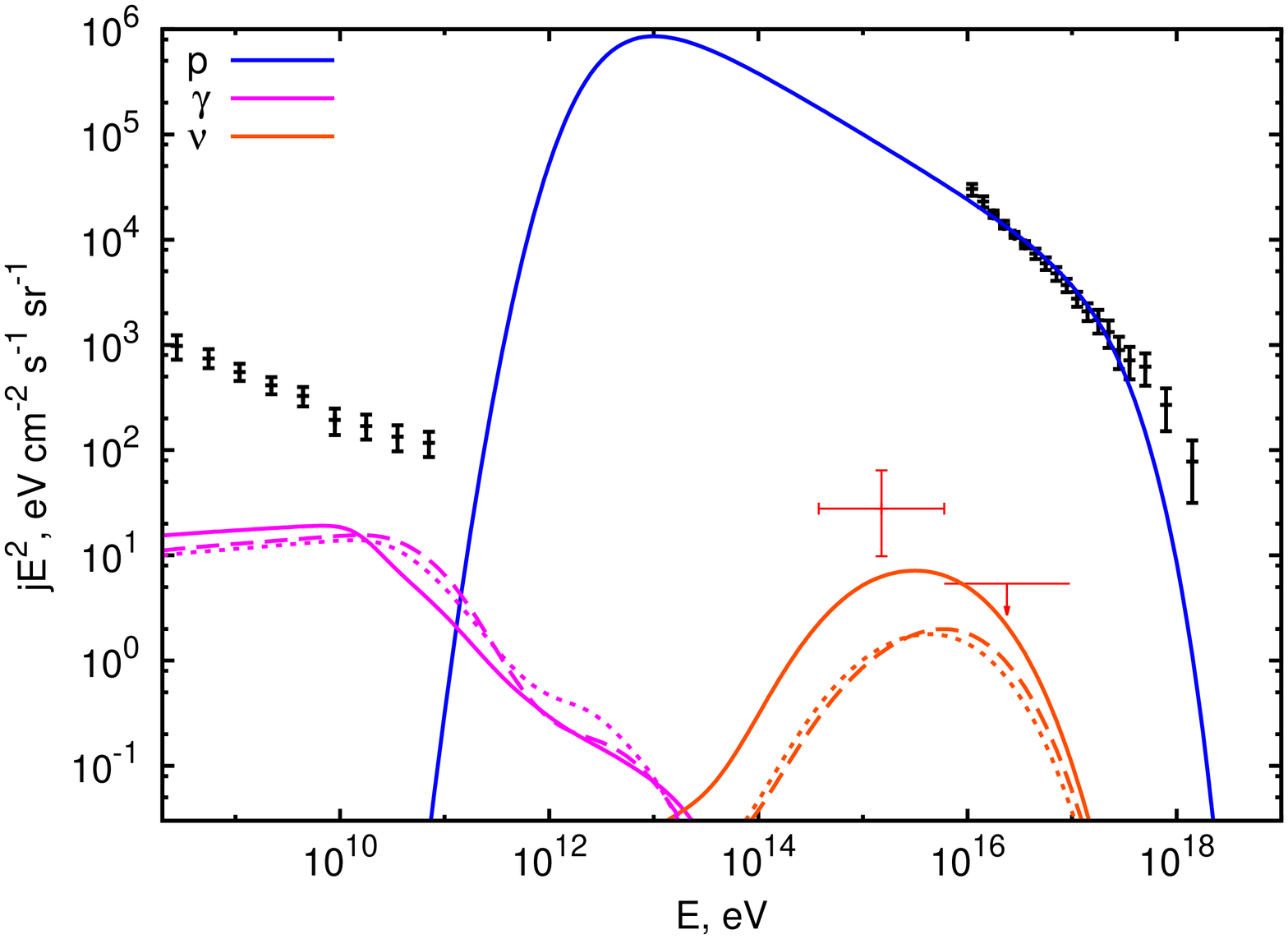}}
\subfigure[]{\includegraphics[angle=270,width=0.47\textwidth]{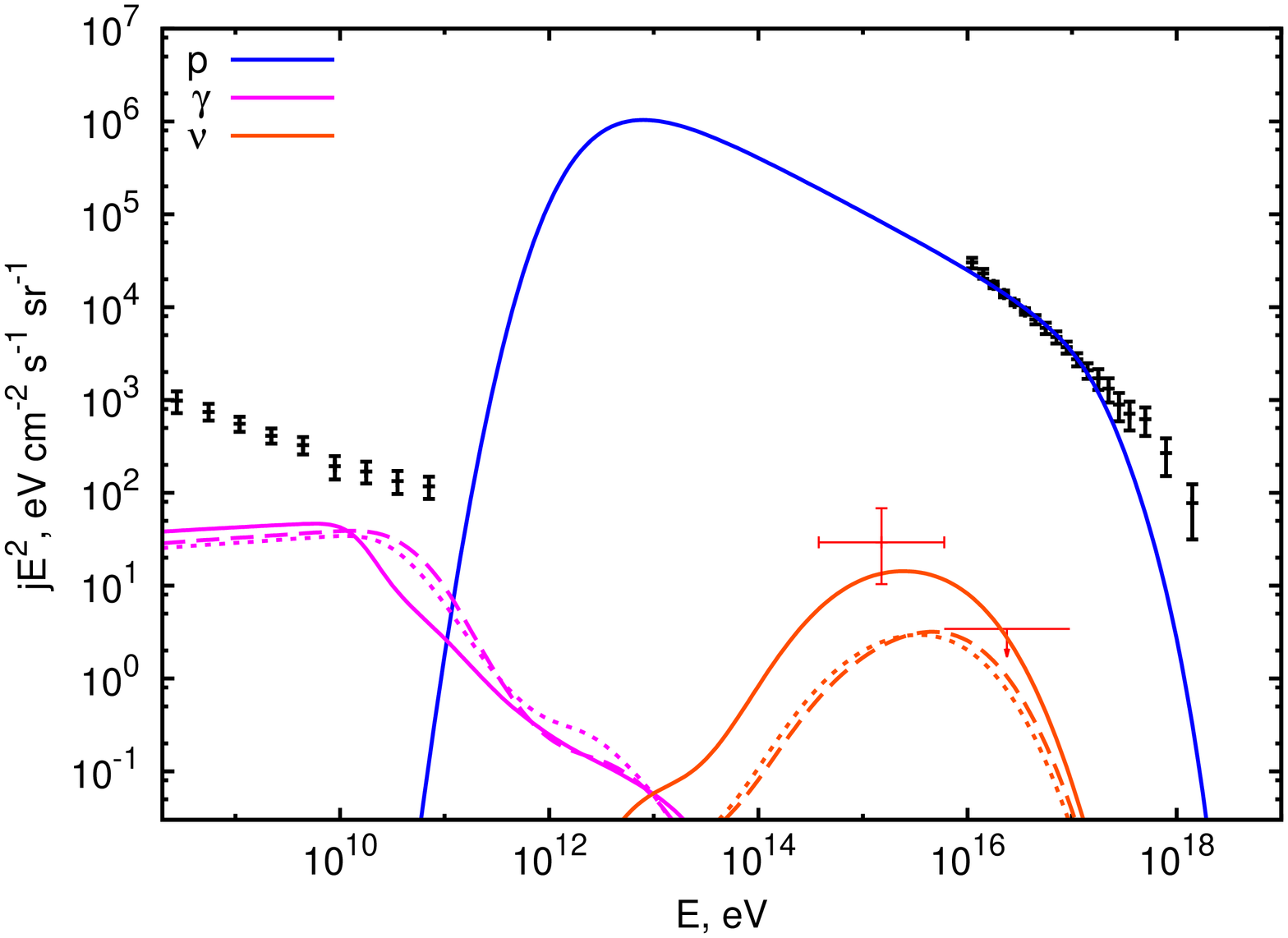}} 
\subfigure[]{\includegraphics[angle=270,width=0.47\textwidth]{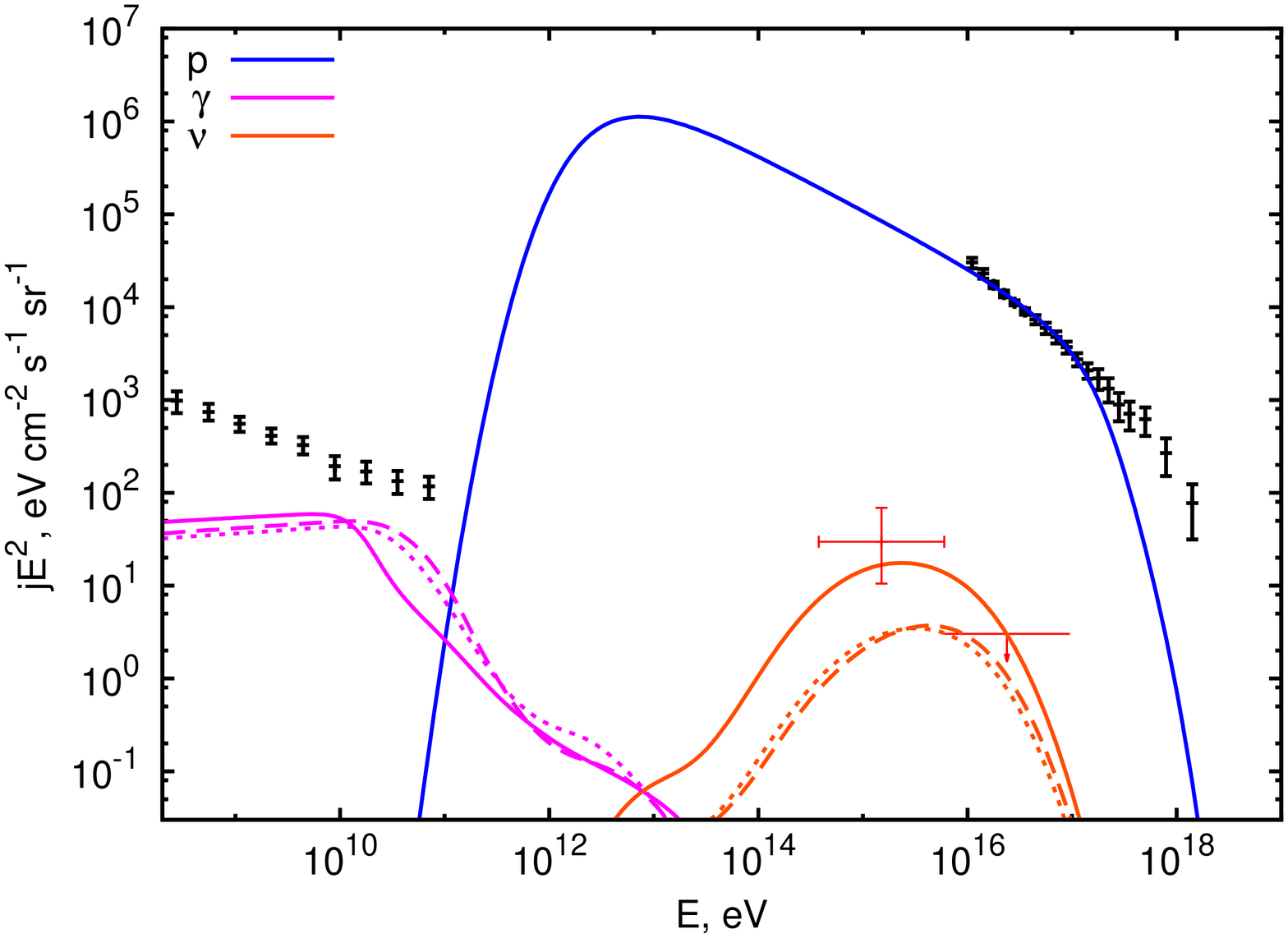}}
\caption{Predicted spectra of PeV neutrinos (red lines) compared with the flux measured by the IceCube experiment~\cite{icecube}. 
The IceCube data points (red) are model-dependent 68\% confidence level flux estimates obtained 
by convolving the IceCube exposure with the predicted neutrino spectrum.  The predicted spectra are shown for the sum of three flavors; each flavor contributes, roughly, 1/3. 
The solid and dotted red lines correspond to the EBL models of Ref.~\cite{Stecker:2005qs} and Ref.~\cite{Inoue:2012bk}, respectively.  
The dashed line represents 
two other models, Refs.~\cite{Kneiske:2003tx} and \cite{Stecker:2012ta}, which yield practically identical spectra. 
The evolution parameters for each plot are listed in Table~\ref{tab:models} for 
(a)  $L_{\rm x}=10^{42.5}$~erg/s, (b) $L_{\rm x}=10^{43.5}$~erg/s, (c)  $L_{\rm x}=10^{44.5}$~erg/s, (d)  $L_{\rm x}=10^{45.5}$~erg/s. 
In all cases, we assumed the proton spectral index $\alpha=2.6$ and the maximal proton energy $E_{\rm p,max}=3\times 10^{17}$~eV.  
Also shown are the predicted gamma ray (lower curves below 10~TeV) and cosmic ray (upper curve) fluxes. 
The cosmic ray data points above 10~PeV are based on KASCADE-Grande~\cite{Apel:2012rm}; the diffuse gamma-ray background data points below 1~TeV are due to Fermi~\cite{Abdo:2010nz}.  
  \label{fig:fourplots}}
\end{figure*}

 A number of different models have been advanced for EBL~\cite{Stecker:2005qs,Kneiske:2003tx,EBL,Stecker:2012ta,Inoue:2012bk}. There are some upper bounds on EBL in the literature that were based on observations of distant blazars, which were derived without taking into account the cosmic ray contribution.  When the cosmic rays are included, these bounds on EBL are relaxed~\cite{Essey:2010er}, and only  the limits based on GRBs~\cite{Abdo:2010kz} remain unaffected.  Based on the photons from GRB 090902B and GRB 080916C observed in the first year of Fermi, one can disfavor the model of Ref.~\cite{Stecker:2005qs} at ``more than 3$\sigma$ level''.  It would be interesting to see an updated analysis of this upper bound based on the much larger dataset available now, after several years of Fermi operations, where one should expect many more gamma rays coincidental with GRBs.  We will consider a broad range of EBL models, including those that show tension with the GRB limit.  Since most the neutrinos are produced 
near the threshold, only the height of the EBL peak  near $1\, \mu$m affects the results.  At those wavelengths, the model of Ref.~\cite{Stecker:2005qs} predicts a higher photon density than most other models.  At the lower side of the range for $1\, \mu$m EBL density are the models of Refs.~\cite{Kneiske:2003tx,Stecker:2012ta,Inoue:2012bk}. We show the spectra for these three models in Fig.~\ref{fig:fourplots}. 

While AGNs are widely expected to accelerate cosmic rays, little is known about the spectrum of cosmic rays produced by a 
typical AGN.  We assume the following form of the proton spectrum: 
\begin{equation}
 j_p(E) \propto E^{-\alpha} \ \exp(-E/E_{\rm p,max}) \, \exp(-E_{\rm p,min}/E).
\end{equation}
The results do not depend on the lower energy cutoff, but the required 
source power does:
\begin{equation}
W \propto \int_{m_p}^{\infty} E j_p(E) dE \simeq \int_{E_{\rm 
p,min}}^{E_{\rm p,max}} E^{1-\alpha} dE \propto E_{\rm p,min}^{2-\alpha},
\end{equation}
for $E_{\rm p,min}/E_{\rm p,max} \ll 1$ and $\alpha>2$.
The lower cutoff in the energy spectrum may exist due to capture of low 
energy protons by the local magnetic fields in the source. Energy 
requirements and the spectral slope of cosmic rays are discussed, e.g., 
in Refs.~\cite{Berezinsky:2002nc,Gaisser:2013bla}.  We used $E_{\rm 
p,min}=10^{13}$~eV, and we explored different values of $E_{\rm p,max}$ 
and $\alpha$.  The best fit to the IceCube flux (without overshooting the diffuse 
cosmic-ray and gamma-ray backgrounds) was obtained for
$\alpha=2.6$, $E_{\rm p,max}=3\times 10^{17}$~eV.  We note that the 
corresponding gamma factor of a proton at the site of acceleration is 
close to the maximal value obtained in some detailed 
simulations~\cite{Sironi:2010rb}. The source power density given below 
in Table~\ref{tab:models} was obtained for $E_{\rm p,min}=10^{13}$~eV. 
If one does not impose a limit on $E_{\rm p,min}$, the power density 
would grow by factor of $(10^{13}{\, \rm eV}/m_p)^{0.6} \simeq 10^{2.4}$.


The contribution of distant sources depends on their evolution with redshift.  Following Ref.~\cite{Hasinger:2005sb}, we parameterize the source density evolution as 
\begin{equation}
 \rho(z) =\left \{
\begin{array}{ll}
 (1+z)^m, & 0<z<z_1 \\
(1+z_1)^m, & z_1<z<z_2 \\
(1+z_1)^m \ 10^{k(z-z_2)}, & z>z_2 
\end{array} \right.
\end{equation}
Here $m, z_1, z_2$, and $k$ are parameters obtained from fitting the observational data; they take different values for different AGN X-ray luminosities $L_{\rm x}$.  
From observational data, Hasinger {\em et al.}~\cite{Hasinger:2005sb} obtain the parameters shown in Table~\ref{tab:models}. We will consider all of these types of redshift evolution 
because one does not know whether the X-ray luminosity is well correlated with the power of cosmic ray emission.

\begin{table}[ht!]
     \begin{tabular}{l|cccc}
        \hline
         \noalign{\smallskip}
        $ L_{\rm x},      {\rm erg/s}  $             & $10^{42.5}$   & $10^{43.5} $  & $ 10^{44.5}$  & $10^{45.5}$   \\
         \noalign{\smallskip}
        \hline
         \noalign{\smallskip}
         $m  $   & $4.0 \pm 0.7$ & $3.4 \pm 0.5$ & $5.0 \pm 0.2$ &$ 7.1 \pm 1.0$ \\  
         $z_1 $                       & $  0.7$  & $  1.2 $ & $  1.7$  &  $ 1.7 $\\  
         $z_2  $                      &  $ 0.7 $ &  $ 1.2 $ & $  2.7 $ &  $ 2.7 $\\  
         $k  $                        &  -0.32 &  -0.32 &  -0.43 &  -0.43\\  
         \noalign{\smallskip}
        \hline
         \noalign{\smallskip}
$W_p, 10^{40} \frac{\rm erg}{{\rm s\, Mpc}^3} \, $   & {7.0} &  {6.0}  &  {1.3}  &  {0.22} \\
         \noalign{\smallskip}
        \hline
     \end{tabular}
\caption{Evolution parameters for AGN with different values of the X-ray power $L_{\rm x}$ inferred from observational data~\cite{Hasinger:2005sb} are shown in the upper part of the table.     
The required power per unit volume $W_p$ of cosmic rays with energies $E_p>10^{13}\, {\rm eV}$  was calculated under the assumption that an average AGN is described by one of these evolution models.
\label{tab:models}}
\end{table}

For each neutrino flavor we calculate the expected number of events in the energy interval of interest by convolving their predicted spectrum with the experimental exposure given in Ref.~\cite{icecube}. The overall flux normalization is chosen on the basis of the following criteria:  (i) the predicted average total number of neutrino events $\bar{N}_{\nu}$ in the energy range $0.4 \, {\rm PeV} < E < 6 \, {\rm PeV}$  must be as close as possible to the observed value ${N}_{\nu}=2$ (68\% CL interval around $2$ is shown in Fig.~\ref{fig:fourplots}); (ii) the Poisson probability to observe at least 1 event above {6 PeV} in the model must be less than 0.68, that is $\bar{N}_{\nu}^{up}<1.14$; (iii) diffuse photon flux should not exceed the {Fermi} upper bound; (iv) the predicted cosmic ray flux should not exceed the observed flux, for which we use the KASCADE-Grande results~\cite{Apel:2012rm}.  In practice, we maximize the PeV neutrino signal without violating any observational constraints. 

The results of our numerical calculations are shown in Fig.~\ref{fig:fourplots}.  As one can see, neutrinos produced in interactions of cosmic rays with background photons can account for the observed neutrino flux reported by IceCube collaboration in the case of strong evolution and high EBL~\cite{Stecker:2005qs}.

The energy requirements per source are consistent with what is expected from AGN.  For each of the models shown in Table~\ref{tab:models} and in  Fig.~\ref{fig:fourplots}, we calculated the emissivity at $z=0$ in cosmic rays with energies above $E_{\rm p,min} = 10^{13}$~eV.  The results vary from $2\times 10^{39}$~erg/s/Mpc$^3$ to $7\times 10^{40}$~erg/s/Mpc$^3$.  Assuming the AGN density of $10^{-5}/$Mpc$^3$~\cite{Treister:2009kw}, one obtains an individual AGN luminosity of $L_0 \simeq 10^{44}$erg/s for the lower end of the above range.  This is a reasonable luminosity, which corresponds to the Eddington mass of $10^6\, M_\odot$.  (AGN jets can exceed the Eddington limit, but, in our case, the average AGN luminosity well below the Eddington luminosity.)  This is also consistent with the analyses of Refs.~\cite{Essey:2010er,Razzaque:2011jc}. 

Future results from IceCube may help constrain models of cosmic ray acceleration in AGN. We note that cosmic ray flux provides a stronger constraint than the diffuse gamma-ray background. 
Composition measurements based on the data of KASCADE-Grande~\cite{Apel:2012rm} are subject to large uncertainties in the Monte Carlo simulations, especially in the energy range of interest to us.
Furthermore, local galactic magnetic fields can affect the flux and composition of cosmic rays with energies below $10^{17}$~eV (and even those with higher energies~\cite{Gaisser:2013bla,UHECR}),
making it difficult to connect the locally measured composition to that of extragalactic sources.  Therefore, we used the total cosmic ray flux as the upper bound. 

In summary, we have examined the recent observations of the IceCube experiment in light of the model that explains the spectra of distant blazars by secondary gamma rays produced in cosmic-ray interactions along the line of sight~\cite{Essey:2009zg,Essey:2009ju,Essey:2010er,Murase:2011cy,Razzaque:2011jc,Essey:2011wv,Prosekin:2012ne,Aharonian:2012fu,Zheng}.  We have shown that he same interactions result in a neutrino spectrum that can be consistent with the IceCube results.

The authors thank J.~Beacom, F.~Halzen, D.~Hooper, M.~Malkan, S.~Scully, L.~Sironi, A.~Spitkovsky, and F.~Stecker for helpful, stimulating discussions. A.K. was supported by DOE Grant DE-FG03-91ER40662 and by the World Premier International Research Center Initiative (WPI Initiative), MEXT, Japan. O.K. was supported by the grant of the Russian Ministry of Education and Science No. 8412  and grant of the President of the Russian Federation NS-5590.2012.2.


\begin{thebibliography}{99}

\bibitem{icecube}
A.~Ishihara, {\em Neutrino 2012},
June, 2012,  Kyoto, Japan;
F.~Halzen,  {\em Neutrino Oscillations Workshop},
September 9--16, 2012, Otranto, Lecce, Italy;
  M.~G.~Aartsen {\it et al.} [IceCube Collaboration],
  arXiv:1304.5356.

\bibitem{Laha:2013lka} 
  R.~Laha, et al., 
  arXiv:1306.2309.

\bibitem{Bhattacharya:2011qu}
A.~Bhattacharya, et al., 
JCAP {\bf 1110} (2011) 017.

\bibitem{Cholis:2012kq}
I.~Cholis and D.~Hooper,
arXiv:1211.1974; 
R.~-Y.~Liu and X.~-Y.~Wang,
Astrophys.\ J.\  {\bf 766}, 73 (2013);
  M.~D.~Kistler, T.~Stanev and H.~Yuksel,
  arXiv:1301.1703.

\bibitem{Essey:2009zg} 
  W.~Essey and A.~Kusenko,
  Astropart.\ Phys.\  {\bf 33}, 81 (2010).
\bibitem{Essey:2009ju}
W.~Essey et al., 
Phys.\ Rev.\ Lett.\  {\bf 104} (2010) 141102.

\bibitem{Essey:2010er} 
  W.~Essey, O.~Kalashev, A.~Kusenko and J.~F.~Beacom,
  Astrophys.\ J.\  {\bf 731}, 51 (2011).
\bibitem{Essey:2011wv} 
  W.~Essey and A.~Kusenko,
  Astrophys.\ J.\  {\bf 751}, L11 (2012).
 
\bibitem{Murase:2011cy} 
  K.~Murase, C.~D.~Dermer, H.~Takami and G.~Migliori,
  Astrophys.\ J.\  {\bf 749}, 63 (2012).
 \bibitem{Razzaque:2011jc} 
  S.~Razzaque, C.~D.~Dermer and J.~D.~Finke,
  Astrophys.\ J.\  {\bf 745}, 196 (2012).
 \bibitem{Prosekin:2012ne} 
  A.~Prosekin, W.~Essey, A.~Kusenko and F.~Aharonian,
  Astrophys.\ J.\  {\bf 757}, 183 (2012).
 \bibitem{Aharonian:2012fu} 
  F.~Aharonian, W.~Essey, A.~Kusenko and A.~Prosekin,
  Phys.\  Rev.\  D87, {\bf 063002} (2013).

\bibitem{Zheng}
Y.~G.~Zheng, T.~ Kang, Astrophys. J. {\bf 764}, 113 (2013).

\bibitem{Essey:2010nd} 
  W.~Essey, S.~Ando and A.~Kusenko,
  Astropart.\ Phys.\  {\bf 35}, 135 (2011).

\bibitem{Stecker:2007zj} 
  F.~W.~Stecker, M.~G.~Baring and E.~J.~Summerlin,
  Astrophys.\ J.\  {\bf 667}, L29 (2007).

\bibitem{Lefa:2011xh} 
  E.~Lefa, F.~M.~Rieger and F.~Aharonian,
  Astrophys.\ J.\  {\bf 740}, 64 (2011).

\bibitem{Dermer:2011uu} 
  C.~Dermer and B.~Lott,
  J.\ Phys.\ Conf.\ Ser.\  {\bf 355}, 012010 (2012).

\bibitem{De_Angelis:2007dy}
  A.~De Angelis, O.~Mansutti and M.~Roncadelli,
  Phys.\ Rev.\  D {\bf 76}, 121301 (2007); 
  M.~Simet, D.~Hooper and P.~D.~Serpico,
  Phys.\ Rev.\  D {\bf 77}, 063001 (2008); 
  D.~Horns et al., 
  Phys.\ Rev.\ D {\bf 86}, 075024 (2012); 
  M.~Meyer, D.~Horns and M.~Raue,
  arXiv:1211.6405. 

\bibitem{Sironi:2010rb} 
  L.~Sironi and A.~Spitkovsky,
  Astrophys.\ J.\  {\bf 726}, 75 (2011);
  L.~Sironi, A.~Spitkovsky and J.~Arons,
  arXiv:1301.5333.

\bibitem{Berezinsky:2002nc} 
  V.~Berezinsky, A.~Z.~Gazizov and S.~I.~Grigorieva,
  Phys.\ Rev.\ D {\bf 74}, 043005 (2006). 

\bibitem{Gaisser:2013bla} 
  T.~K.~Gaisser, T.~Stanev and S.~Tilav,
  arXiv:1303.3565.

\bibitem{UHECR}
  R.~Aloisio, V.~Berezinsky and A.~Gazizov,
  Astropart.\ Phys.\  {\bf 34}, 620 (2011); 
  A.~Calvez, A.~Kusenko and S.~Nagataki,
  Phys.\ Rev.\ Lett.\  {\bf 105}, 091101 (2010).

\bibitem{Roulet:2012rv} 
  E.~Roulet, G.~Sigl, A.~van Vliet and S.~Mollerach,
 JCAP {\bf 1301}, 028 (2013).

\bibitem{Kistler:2013my} 
  M.~D.~Kistler, T.~Stanev and H.~Yuksel,
  arXiv:1301.1703.

\bibitem{Gelmini:2011kg} 
  G.~B.~Gelmini, O.~Kalashev and D.~V.~Semikoz,
  JCAP {\bf 1201}, 044 (2012).
  
\bibitem{Stecker:2005qs} 
  F.~W.~Stecker, M.~A.~Malkan and S.~T.~Scully,
  Astrophys.\ J.\  {\bf 648}, 774 (2006).


\bibitem{Kneiske:2003tx} 
T.~M.~Kneiske et al., 
Astron.\ Astrophys.\  {\bf 386}(2002) 1; {\em ibid.},  
{\bf 413} (2004) 807.


\bibitem{EBL}
  J.~R.~Primack, R.~C.~Gilmore and R.~S.~Somerville,
  AIP Conf.\ Proc.\  {\bf 1085}, 71 (2009);
  A.~Franceschini, G.~Rodighiero and M.~Vaccari,
  Astron.\ Astrophys.\  {\bf 487}, 837 (2008);
  J.~D.~Finke, S.~Razzaque and C.~D.~Dermer,
  Astrophys.\ J.\  {\bf 712}, 238 (2010).

\bibitem{Stecker:2012ta} 
  F.~W.~Stecker, M.~A.~Malkan and S.~T.~Scully,
  Astrophys.\ J.\  {\bf 761}, 128 (2012).

 \bibitem{Inoue:2012bk} 
  Y.~Inoue et al., 
  arXiv:1212.1683. 

\bibitem{Abdo:2010kz} 
  A.~A.~Abdo {\it et al.} 
  Astrophys.\ J.\  {\bf 723}, 1082 (2010).

\bibitem{Hasinger:2005sb} 
  G.~Hasinger, T.~Miyaji and M.~Schmidt,
  Astron.\ Astrophys.\  {\bf 441}, 417 (2005).

\bibitem{Apel:2012rm} 
W.~D.~Apel {\it et al.} 
arXiv:1206.3834.
   
\bibitem{Abdo:2010nz} 
  A.~A.~Abdo {\it et al.} 
  Phys.\ Rev.\ Lett.\  {\bf 104}, 101101 (2010). 
 
\bibitem{Treister:2009kw} 
  E.~Treister et al.,
  Astrophys.\ J.\  {\bf 696}, 110 (2009).

\end{thebibliography}
\end{document}